\documentclass[useAMS,usenatbib]{mn2e}

\usepackage{graphicx}
\usepackage{graphics}
\usepackage{mathtools}
\usepackage{url}
\usepackage{hyperref}

\newcommand{\kbol}{k_\mathrm{B}}
\newcommand{\dd}{\mathrm{d}}

\begin{document}

\title[Electron-attachment rates for carbon-rich molecules]{Electron-attachment rates for carbon-rich molecules in protoplanetary atmospheres: the role of chemical differences}

\author[F. Carelli et al.]{F. Carelli$^1$, 
	T. Grassi$^1$, F. Sebastianelli$^1$,  and F. A. Gianturco$^1$\thanks{Corresponding author: e-mail: fa.gianturco@caspur.it Fax +39-06-49913305}\\
$^{1}$Dept. of Chemistry, University of Rome ``Sapienza'', P.le A. Moro 5, 00161 Rome, Italy\\}

\date{Accepted *****. Received *****; in original form ******}

\pagerange{\pageref{firstpage}--\pageref{lastpage}} \pubyear{2012}

\maketitle
\label{firstpage}

\begin{abstract}
The formation of anionic species in the interstellar medium from interaction of linear molecules containing carbon, nitrogen and hydrogen as atomic components (polyynes) with free electrons in the environment is modelled via a quantum treatment of the collision dynamics. The ensuing integral cross sections are employed to obtain the corresponding attachment rates over a broad range of temperatures for the electrons. The calculations unequivocally show that a parametrization form often employed for such rates yields a broad range of values that turn out to be specific for each molecular species considered, thus excluding using a unique set for the whole class of polyynes.
\end{abstract}


\section{Introduction}
In spite of the fact that the physical conditions in the Interstellar Medium (ISM) and in many of the planetary atmospheres are seemingly unfavourable to chemical evolutions (e.g. low particle densities and fairly low temperatures) a very rich chemistry exists in such regions and has been followed by observations for a long time \citep{Snow2006, Herbst2009}. Among the recently discovered species, six different anionic molecules have been detected in the denser regions of the ISM, i.e. HC$_\mathrm{n}^-$ (n$=4,6,8$) and  C$_\mathrm{n}$N$^-$ (n$=1,3,5$) \citep{McCarthy2006, Brunken2007,Cernicharo2007,Agundez2010} and furthermore N-containing anions C$_\mathrm{n}$N$^-$ (n$=1,3,5$) were found in Titan's upper atmosphere \citep{Vuitton2009}. It is therefore well established by now that several molecular anions have been found in the interstellar and planetary environments, so that it is now crucial to understand as much as possible the ion chemistry of the species involved and the likely paths to the formation of such systems: the final rate constants for their formation are in fact important for the astrochemical modelling that tries to establish the most realistic reaction networks in that environment.

Given the low densities of the mutual species present in the ISM, it is also a realistic possibility that such anionic molecules be formed in reaction with electrons, the latter being produced abundantly in the outer regions of the denser clouds and in the planetary atmospheres \citep{Sakai2007}. Estimates of the electron temperatures in that environment suggest values that can go up to 1000 K or above.
The reactions of such electrons with molecular species can therefore lead to the eventual formation of a negative ion through a variety of dynamical processes. For example, dissociative electron attachment (DEA) is a well known process
\begin{equation}
	\mathrm{M\!-\!H}+\mathrm{e}^-\to\mathrm{M}^-+\mathrm{H}\,,
\end{equation}
whereby some of the bound atoms in the initial partner are detached after forming a stable residual anion, a process related to the sign and the value of the electron affinity (EA) of the $\mathrm{M\!-\!H}$ species \citep{Sebastianelli2011}. Another option in the ISM environment is the radiative stabilization (RS) of the threshold resonances associated to the formation of a temporary negative ion (TNI) upon electron attachment close to zero
energy
\begin{equation}
	\mathrm{M}+\mathrm{e}^-\to\left[\mathrm{M}^-\right]^*\to\mathrm{M}^-+h\nu
\end{equation}
a process which we have recently argued as possible for the formation of $\left[\mathrm{C_6H_4}\right]^-$ \citep{Carelli2011}.
The DEA processes are also known to be more likely to occur at low energies and to chiefly involve a dehydrogenation path \citep{Graupner2006}
\begin{equation}
	\mathrm{e}^-+\mathrm{HCCCN}\to\mathrm{CCCN}^-+\mathrm{H}
\end{equation}
which has, in fact, been suggested by our calculations as occurring via resonant intermediates in several polyynes anions \citep{Sebastianelli2012}. An additional feature that should favor RS in competition with DEA is the occurrence of virtual states and of zero-energy resonances in unsaturated species like C$_{60}$ \citep{Lucchese1999}. Such special ``compound states'' are suggested to be very efficient in ``soaking up'' threshold electrons, thereby leading to anionic formation \citep{Field2001} and have been found by our calculations to also exist in anionic polyynes \citep{Sebastianelli2012}.

\section{Calculating attachment rates}
The question of having such variety of processes thus relates to scattering cross sections over the full range of low-energy regions where either DEA and RS processes are found to occur. They often reveal a broad variety of resonances for metastable electron attachment \citep{Sebastianelli2012}. Such cross sections in turn lead to the corresponding rates given as
\begin{equation}\label{ratecalc}
	k_\mathrm{TNI}(T) = \left(\frac{8\kbol T}{\pi\mu_\mathrm{e}}\right)^{1/2}\frac{1}{(\kbol T)^2}
		\int E\sigma_0(E) e^{-E/\kbol T}\dd E\,.
\end{equation} 
In our present models it describes a convolution over electron temperatures of the integral, total cross sections $\sigma_0(E)$ that include all metastable attachment processes for the given molecule over that range of energies, plus a sum over elastic and rotational excitation processes starting with $|j\rangle=0$ states of the targets. In our procedure for the calculation of the relevant cross sections - e.g. see for details our foundation paper: \citet{Lucchese1996} - the channels associated with electronic excitations and vibrational excitations have been excluded to better model ISM conditions, although the entire quantum dynamics of the scattering of electrons from the polyynes has been carried out using \emph{ab initio} methods - see again: \citet{Carelli2011}. We also know, however, from previous analysis of polyatomic targets, that the inelastic cross sections contribute, at the relevant energy, about 10\% of the integral cross sections included in 
Eqn.(\ref{ratecalc}) - \citet{Irrera2005}.

One should also note that the $\sigma_0$ computed cross sections of Eqn.(\ref{ratecalc}) are obtained for fixed nuclear geometries of the target polyynes and therefore the needed nuclear dynamical couplings with the electron motion, necessary to obtain DEA cross sections, are not included. However, they include direct and accurate evaluations from first principles of all the resonance formation states and all the virtual state effects. Thus, as we shall further explain below, the corresponding rates would be upper bound to the true rates since all integral cross sections are taken to lead the anionic stabilization.

The whole electron attachment process, namely
\begin{equation}\label{kall}
	\mathrm{M}(R_i) + \mathrm{e}^-\xrightarrow{k_\mathrm{EA}}\mathrm{M}(R_f)^-
\end{equation}
can be divided into the primary attachment (labelled with $k_\mathrm{TNI}$) and a subsequent process ($k_\alpha$) that, via intramolecular vibrational rearrangements, can lead to a stable anion according to the following prototypical reaction:
\begin{equation}
	\mathrm{M}(R_i) + \mathrm{e}^-\xrightarrow{k_\mathrm{TNI}} \left[\mathrm{M}^-\right]^*\xrightarrow{k_\alpha}\mathrm{M}(R_f)^-\,.
\end{equation}
In the above equation $k_\alpha$ refers either to the radiative stabilization ($\alpha=\mathrm{rs}$) or 
to the dissociative process ($\alpha=\mathrm{DEA}$), i.e.
\begin{equation}
	\left[\mathrm{M}^- (R_e)\right]^*\xrightarrow{k_\mathrm{rs}}\mathrm{M}(R_f)^- + h\nu
\end{equation}
and
\begin{equation}
	\left[\mathrm{M}^-(R_e)\right]^*\xrightarrow{k_\mathrm{DEA}}\left[\mathrm{M}-\mathrm{A}\right]^- + \mathrm{A}\,,
\end{equation}
both in competition with the auto-detachment process
\begin{equation}
	\mathrm{M}(R_e) + \mathrm{e}^- \to \left[\mathrm{M}^-\right]^*\xrightarrow{k_\mathrm{ad}}\mathrm{M}(R_f)+\mathrm{e}^-\,.
\end{equation}
In the above equations the changes in molecular initial geometries, $R_i$, are indicated by the final geometries, $R_f$, or by the formation of an anion in its initial equilibrium geometry, $R_e$. Note also that DEA forms a different molecular species $\left[\mathrm{M}-\mathrm{A}\right]^-$.

All the above processes are taken to occur via the formation of resonance states, transient negative ions (TNI) that play the role of gateway states
to the final formation of all stable anions, as described in Eqn.(\ref{kall}).

Thus, the present calculations generate from first principle quantum-dynamics - see \citet{Lucchese1996} for all details - all the TNI states contributing
to the rate of Eqn.(\ref{kall}), plus all the zero-energy resonances and virtual states that occur as the collision energy goes down to zero
and which can further contribute to $k_\mathrm{EA}$.

One should also be reminded that the overall rate of electron attachment (EA), $k_\mathrm{EA}$, can be related to the rates of metastable anion formations, $k_\mathrm{TNI}$, and to the autodetachment and radiative stabilization rates, $k_\mathrm{ad}$
and $k_\mathrm{rs}$ respectively \citep{CarelliPhdThesis}
\begin{equation}\label{eqnEA}
	k_\mathrm{EA}=\frac{ k_\mathrm{rs}}{k_\mathrm{ad}+k_\mathrm{rs}}k_\mathrm{TNI}
\end{equation}
when anionic formation is assumed to be in a steady state \citep{Herbst1980}. Since it is also assumed that $k_\mathrm{ad}\gg k_\mathrm{rs}$ one can further write
\begin{equation}
	k_\mathrm{EA}=\frac{k_\mathrm{rs}}{k_\mathrm{ad}} k_\mathrm{TNI}\,.
\end{equation}
Therefore the rates of electron attachment, $k_\mathrm{EA}$, have a clear upper bound in the calculated $k_\mathrm{TNI}$ rates which use Eqn.(\ref{ratecalc}) unless
$k_\mathrm{ad}\approx k_\mathrm{rs}$.
The present calculations for polyynes are thus providing from first principles electron attachment rate values which are, generally speaking, larger than the true rates of formation of anionic species.
They are also shown here to exhibit a very marked dependence on the chemical features of the carbon-rich linear molecules examined in the present work.

\section{Results and Discussion}
We have carried out detailed calculations of the integral cross sections, as in Eqn.(\ref{ratecalc}), over
a broad range of collision energies using the computational method well established in our group \citep{Lucchese1996}
as already given in detail in various papers on polyyne species - e.g. see: \citet{Sebastianelli2012, Sebastianellietal2012}.

It is also customary in evolutionary models of chemical networks in the ISM and in planetary atmospheres to represent general two-body reactions via a parametric form of their corresponding rate coefficients - e.g. see: \citet{Woodall2007}
\begin{equation}\label{abcfit}
	k_\mathrm{2B}=\alpha\left(\frac{T}{300\,\mathrm{K}}\right)^\beta\exp\left(-\gamma/T\right)\,\mathrm{cm}^3\mathrm{s}^{-1}\,,
\end{equation}
where, in our case, $T$ is the electron temperature $T_e$. The present calculations can therefore be profitably represented by a set of three parameters: $\alpha, \beta, \gamma$, that can be associated to each polyyne species
considered within the chemical network.

\begin{table}
	\caption{Fit coefficients for the polyynes presented in this paper, where 
		$\alpha$ is in cm$^3$ s$^{-1}$, $\beta$ is dimensionless, and $\gamma$ in K. Note that $a(b)=a\times10^b$. See text and Eqn.(\ref{abcfit}) for further details.}
	\begin{center}
		\begin{tabular}{l|rrr}
			\hline
			 & $\alpha$ & $\beta$ & $\gamma$\\
			\hline
			HC$_4$H & 2.689(-10) & 0.486 & 4.174\\
			NC$_2$N & 1.887(-10) & 0.373 & -5.440\\
			NC$_4$N & 21.798(-10) & -0.019 & 7.688\\
			HC$_3$N & 3.321(-8) & -0.304 & 19.036\\
			NC$_5$N & 7.218(-8) & -0.482 & 7.625\\
			\hline
		\end{tabular}
	\end{center}
	\label{tab_abc}
\end{table}

The molecules which we consider correspond to primary molecules for which electron attachment rates can be calculated: the symmetric polyynes (HC$_4$H, NC$_2$N, and NC$_4$N) and the polar polyynes (HC$_3$N and HC$_5$N)
as exemplary choices for the class of molecules which have been also experimentally observed (see present introduction).

\begin{figure}
\includegraphics[width=.45\textwidth]{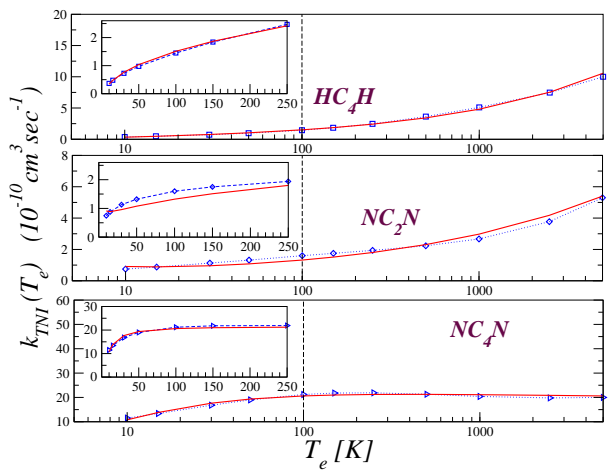} 
\caption{Computed $k_\mathrm{TNI}$ rates from Eqn.(\ref{ratecalc}) for three different carbon-rich linear molecules without 
	permanent dipoles. See text for further details.\label{kf_nodipole}}
\end{figure}

\begin{figure}
\includegraphics[width=.45\textwidth]{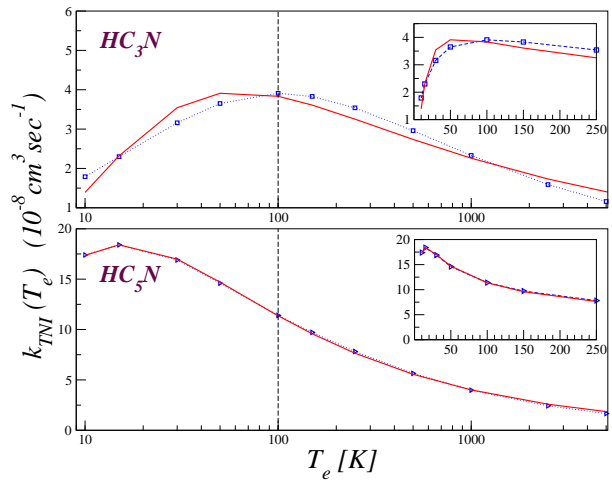} 
\caption{Same as Fig.\ref{kf_nodipole} but for two polar polyynes.\label{kf_dipole}}
\end{figure}

The data reported by Fig.\ref{kf_nodipole} and Fig.\ref{kf_dipole} show the temperature dependence of the $k_\mathrm{TNI}$
from the present calculations over a very broad range of $T_e$: from nearby zero to 5000 K.

The data in both figures  are then fitted using the customary expression of Eqn.(\ref{abcfit}) and the corresponding
parameter values are also reported (see also Tab.\ref{tab_abc}). 

The following features could be gleaned from these data:
\begin{enumerate}
\item There is a dramatic change in size when going from polar to non-polar molecular partners: the rates increase for the former by nearly two orders of magnitude over the whole range of $T_e$;
\item The polar targets also show a strong increase at low $T_e$ values, decreasing at larger temperatures; the opposite occurs for the symmetric polyynes, whose rates increase as $T_e$ increases;
\item The parameters for the polar systems are fairly robust, in the sense that change little with the number of carbon atoms, a molecular index often used in astrochemical networks;
\item The symmetric polyynes show instead a more marked dependence on the number of C atoms and also have larger rates with N atoms in the chains.
\end{enumerate}

The data insets of Fig.\ref{kf_nodipole} and Fig.\ref{kf_dipole} also report the $k_\mathrm{TNI}$ behaviour in the low-$T_e$ regimes,
i.e. from threshold  to 250 K, in order to show more clearly the difference in behaviour among chemical partners.

The marked increase of the rates near threshold and their drops at vanishing energies are again very clear
for the polar molecules, as well as the larger sizes of their rates at low $T$. On the contrary, the symmetric 
polyynes show all a steady increase in size from threshold and overall sizes which are much smaller that those for
their asymmetrical counterparts.

In conclusion, the present quantum calculations of formation rates for electron attachment processes
within a representative sampling of carbon-rich linear molecules show rather clearly that the parametrization
of such two-body processes is markedly dependent on the chemical species being considered. Although the length
of the chains (i.e. the number of C atoms) plays some role, as does the presence of other heteroatoms like
nitrogen, we found here that the strongest chemical discriminator is provided by the presence of permanent electric
dipole moments in the molecules in question. The asymmetric polyynes are seen here to yield much larger rates, especially for $T_e$ values from threshold to about 100 K, and to exhibit an increase at lower temperatures, the opposite
behaviour from that shown by symmetric molecular partners.

Such differences should therefore be taken into consideration when anionic formation rates are explicitly included in evolutionary chemical networks,
where they can provide more detailed, chemically-controlled, final stationary abundances for neutral and charged species. The latter data
would obviously be critical for a better comparison with observations - e.g. see \citet{Agundez2010}.
\section*{Acknowledgements}
The computational support from the CASPUR Consortium is gratefully acknowledged, as well as the financial
support from the PRIN 2009 research network. One of us (T.G.) thanks the CINECA Consortium
for the awarding of a postdoctoral grant during which this work was carried out.

\bibliographystyle{apj}      
\bibliography{mybib} 

\begin{thebibliography}{20}
\expandafter\ifx\csname natexlab\endcsname\relax\def\natexlab#1{#1}\fi

\bibitem[{{Ag{\'u}ndez} {et~al.}(2010){Ag{\'u}ndez}, {Cernicharo},
  {Gu{\'e}lin}, {Kahane}, {Roueff}, {K{\l}os}, {Aoiz}, {Lique}, {Marcelino},
  {Goicoechea}, {Gonz{\'a}lez Garc{\'{\i}}a}, {Gottlieb}, {McCarthy}, \&
  {Thaddeus}}]{Agundez2010}
{Ag{\'u}ndez}, M., {Cernicharo}, J., {Gu{\'e}lin}, M., {et~al.} 2010, \aap,
  517, L2

\bibitem[{{Br{\"u}nken} {et~al.}(2007){Br{\"u}nken}, {Gupta}, {Gottlieb},
  {McCarthy}, \& {Thaddeus}}]{Brunken2007}
{Br{\"u}nken}, S., {Gupta}, H., {Gottlieb}, C.~A., {McCarthy}, M.~C., \&
  {Thaddeus}, P. 2007, \apjl, 664, L43

\bibitem[{{Carelli}(2012)}]{CarelliPhdThesis}
{Carelli}, F. 2012, {``Molecular anions in circumstellar envelopes,
  interstellar clouds, and planetary atmospheres: quantum dynamics of formation
  and evolution''}, \url{http://arxiv.org/abs/1209.2827}

\bibitem[{Carelli {et~al.}(2011)Carelli, Sebastianelli, Satta, \&
  Gianturco}]{Carelli2011}
Carelli, F., Sebastianelli, F., Satta, M., \& Gianturco, F.~A. 2011, Monthly
  Notices of the Royal Astronomical Society, 415, 425

\bibitem[{{Cernicharo} {et~al.}(2007){Cernicharo}, {Gu{\'e}lin}, {Ag{\"u}ndez},
  {Kawaguchi}, {McCarthy}, \& {Thaddeus}}]{Cernicharo2007}
{Cernicharo}, J., {Gu{\'e}lin}, M., {Ag{\"u}ndez}, M., {et~al.} 2007, \aap,
  467, L37

\bibitem[{{Field} {et~al.}(2001){Field}, {Ziesel}, {Lunt}, {Parthasarathy},
  {Suess}, {Hill}, {Dunning}, {Lucchese}, \& {Gianturco}}]{Field2001}
{Field}, D., {Ziesel}, J.-P., {Lunt}, S.~L., {et~al.} 2001, Journal of Physics
  B Atomic Molecular Physics, 34, 4371

\bibitem[{{Graupner} {et~al.}(2006){Graupner}, {Merrigan}, {Field}, {Youngs},
  \& {Marr}}]{Graupner2006}
{Graupner}, K., {Merrigan}, T.~L., {Field}, T.~A., {Youngs}, T.~G.~A., \&
  {Marr}, P.~C. 2006, New Journal of Physics, 8, 117

\bibitem[{{Herbst}(1980)}]{Herbst1980}
{Herbst}, E. 1980, \apj, 237, 462

\bibitem[{{Herbst} \& {van Dishoeck}(2009)}]{Herbst2009}
{Herbst}, E., \& {van Dishoeck}, E.~F. 2009, \araa, 47, 427

\bibitem[{{Irrera} \& {Gianturco}(2005)}]{Irrera2005}
{Irrera}, S., \& {Gianturco}, F.~A. 2005, New Journal of Physics, 7, 1

\bibitem[{Lucchese {et~al.}(1999)Lucchese, Gianturco, \& Sanna}]{Lucchese1999}
Lucchese, R.~R., Gianturco, F., \& Sanna, N. 1999, Chemical Physics Letters,
  305, 413

\bibitem[{{Lucchese} \& {Gianturco}(1996)}]{Lucchese1996}
{Lucchese}, R.~R., \& {Gianturco}, F.~A. 1996, International Reviews in
  Physical Chemistry, 15, 429

\bibitem[{{McCarthy} {et~al.}(2006){McCarthy}, {Gottlieb}, {Gupta}, \&
  {Thaddeus}}]{McCarthy2006}
{McCarthy}, M.~C., {Gottlieb}, C.~A., {Gupta}, H., \& {Thaddeus}, P. 2006,
  \apjl, 652, L141

\bibitem[{{Sakai} {et~al.}(2007){Sakai}, {Sakai}, {Osamura}, \&
  {Yamamoto}}]{Sakai2007}
{Sakai}, N., {Sakai}, T., {Osamura}, Y., \& {Yamamoto}, S. 2007, \apjl, 667,
  L65

\bibitem[{Sebastianelli {et~al.}(2012)Sebastianelli, Carelli, \&
  Gianturco}]{Sebastianellietal2012}
Sebastianelli, F., Carelli, F., \& Gianturco, F. 2012, Chemical Physics, 398,
  199 , chemical Physics of Low-Temperature Plasmas (in honour of Prof Mario
  Capitelli)

\bibitem[{Sebastianelli {et~al.}(2011)Sebastianelli, Carelli, \&
  Gianturco}]{Sebastianelli2011}
Sebastianelli, F., Carelli, F., \& Gianturco, F.~A. 2011, The Journal of
  Physical Chemistry A, 115, 11531

\bibitem[{{Sebastianelli} \& {Gianturco}(2012)}]{Sebastianelli2012}
{Sebastianelli}, F., \& {Gianturco}, F.~A. 2012, European Physical Journal D,
  66, 41

\bibitem[{{Snow} \& {McCall}(2006)}]{Snow2006}
{Snow}, T.~P., \& {McCall}, B.~J. 2006, \araa, 44, 367

\bibitem[{{Vuitton} {et~al.}(2009){Vuitton}, {Lavvas}, {Yelle}, {Galand},
  {Wellbrock}, {Lewis}, {Coates}, \& {Wahlund}}]{Vuitton2009}
{Vuitton}, V., {Lavvas}, P., {Yelle}, R.~V., {et~al.} 2009, \planss, 57, 1558

\bibitem[{{Woodall} {et~al.}(2007){Woodall}, {Ag{\'u}ndez}, {Markwick-Kemper},
  \& {Millar}}]{Woodall2007}
{Woodall}, J., {Ag{\'u}ndez}, M., {Markwick-Kemper}, A.~J., \& {Millar}, T.~J.
  2007, \aap, 466, 1197

\end{thebibliography}

\bsp

\label{lastpage}
\end{document}